\documentstyle[psfig]{aa}    

\newcommand{\tuc}{47$\,$Tuc}

\newcommand{\x}{X$\,$}
\newcommand{\vv}{V$\,$}
\newcommand{\ako}{AKO$\,$}
\newcommand{\ltap}{\mathrel{\hbox{\rlap{\lower.55ex \hbox {$\sim$}}
                   \kern-.3em \raise.4ex \hbox{$<$}}}}
\newcommand{\gtap}{\mathrel{\hbox{\rlap{\lower.55ex \hbox {$\sim$}}
                   \kern-.3em \raise.4ex \hbox{$>$}}}}
\newcommand{\nh}{N_{\rm H}}
\newcommand{\cmsq}{{\rm cm}^{-2}}
\begin{document}
\thesaurus{05(10.07.3, 13.25.5)}
\title{Nine X-ray sources in the globular cluster 47~Tucanae}
\author{F.~ Verbunt\inst{1} \and G.~Hasinger\inst{2} 
 }
\offprints{F.~Verbunt}

\institute{     Astronomical Institute,
              P.O.Box 80000, NL-3508 TA Utrecht, The Netherlands
         \and              
              Astrophysikalisches Institut, An der Sternwarte 16,
              D-14482 Potsdam, Germany
                        }
\date{Received, accepted 25 May 1998}   
\authorrunning{F.\ Verbunt \&\ G. Hasinger}
\maketitle


\begin{abstract}
We analyze ROSAT HRI observations obtained from 1992 to 1996
of the globular cluster \tuc.
Identifications of two X-ray sources with HD~2072 and with a galaxy,
respectively, are used to obtain accurate ($<2''$) positions of the
X-ray sources in the cluster. 
We find possible optical counterparts, including the  blue objects \vv1 
and \vv2, for three X-ray sources in the core of \tuc, but note that
the probability of chance positional coincidence is significant.

One of the five sources previously reported by us to reside in the
cluster core is found to be an artefact of misalignment between
subsequent satellite pointings.
\end{abstract}

\section{Introduction}

The cores of globular clusters harbour many interesting objects
detected at different wavelengths, such as X-ray sources, ultraviolet and
visual variables and blue stragglers, and radio pulsars.
The sheer number density of stars in the cluster cores makes identification
of sources detected in one wavelength range with those found at
other wavelengths a daunting task, especially for X-ray sources
whose positions are uncertain by more than an arcsecond at best.

As an example, X-ray sources detected in the core of 
\tuc\ (Hasinger et al. 1994, henceforth called Paper~1)\nocite{hjv94}
have been tentatively identified with a cataclysmic variable
(Paresce et al.\ 1992), with blue stragglers (Meylan et al.
1996), and with a remarkable ultraviolet variable (Auri\`ere
et al.\ 1989, Minniti et al.\ 1997).
\nocite{pmf92}\nocite{mmp+96}\nocite{ako89}\nocite{mmp+97}
These various options are possible due to the uncertainty in the
absolute positions of the X-ray sources of about 5$\,''$.

In this paper we analyse three new ROSAT HRI observations of
the globular cluster \tuc, and re-analyse two. With use of the
detailed astrometric study by Geffert et al.\ (1997) we try to obtain an
absolute accuracy of the X-ray positions at the arcsecond level.
\nocite{gak97}
In Sect.\ 2 we describe the observations and data analysis, and in Sect.\ 3
the results. A discussion follows in Sect.\ 4.

\section{Observations and data analysis}

The log of the observations with the ROSAT X-ray telescope
(Tr\"umper et al.\ 1991) \nocite{tha+91} 
together with 
the high - resolution imager 
(HRI, David et al.\ 1992) \nocite{dhkz92} is given in Table~\ref{ta}.
For reasons explained below, we treat the
April 1992 and May 1992 data as two separate observations
(in contrast to Paper~1, where these were treated as a single observation).
The data were reduced with the Extended Scientific Analysis System (EXSAS;
Zimmermann et al.\ 1996).\nocite{zbb+96}

\begin{table}
\caption[o]{Log of the ROSAT HRI observations of \tuc. For each separate 
observation, indicated with a-f, the observation date(s), and
exposure time are given. We further give the shift in $''$
applied to bring the observation to the X-ray coordinate frame of c.
Observation e was timed to be quasi-simultaneous with a Hubble Space
Telescope observation of \tuc\ (see Minniti et al.\ 1997).
\label{ta}}
\begin{tabular}{ll@{\hspace{0.2cm}}ll@{\hspace{0.2cm}}rl@{\hspace{0.2cm}}l}
& \multicolumn{3}{c}{\mbox observing period} & $t_{\rm exp}$(s)
& $\Delta_x('')$ & $\Delta_y('')$ \\
a & 1992 & Apr & 2448732.006--32.020 & 1169  & +2.5 & $-$2.0 \\
b & 1992 & May & 2448764.413--65.424 & 3370 & $-$2.0 & +2.5 \\
c & 1993 & Apr & 2449094.509--96.913 & 13247 &  \\
d & 1994 & Nov & 2449686.500--93.616 & 18914 & +1.5 & $-$1.5 \\
e & 1995 & Oct & 2450015.817--16.073 & 4579  & +2.0 & $-$0.5 \\
f & 1996 & Nov & 2450404.079--18.418 & 17542 & +1.5 & $-$1.0 \\
\end{tabular}
\end{table}

The analysis of the new HRI data and reanalysis of the 1992/1993 data
previously discussed in Paper~1, is done in several steps.
First, we use the X-ray positions of four bright sources that are detectable
in the separate observations to align all observations onto a 
single X-ray coordinate frame. We then use two optical identifications to 
convert the X-ray coordinate frame to J2000.
The analysis of the sources is done separately for the region away
from the cluster core, for which standard EXSAS techniques can be applied,
and for the inner region, where we use contour maps and a multiple-source
detection procedure to determine the individual source positions.
The details of these procedures are as follows.

\subsection{Co-alignment of the separate observations}

Each photon is detected at a detector position $(x,y)$.
This position gives the distance to the center of
the HRI detector. The pointing direction of the satellite provides
the celestial coordinates for the center of the detector, and allows
the conversion for each detector position to a celestial position.
For this conversion a pixel size of $0.5''$ is assumed.
The uncertainty in the pointing direction
causes small shifts between the celestial positions derived from
different observations.
When different observations are added such shifts lead to smearing,
and in extreme cases even doubling, of the sources.
Before adding the separate observations, we therefore determine the
shifts between them.

We first take account of the re-determination of the size of the HRI pixels 
to $0.4986''$ (Hasinger et al.\ 1998), \nocite{hbg+98}
by multiplying the $x,y$ pixel coordinates of each photon with $0.9972$.
(This enables us to use the EXSAS software, which assumes $0.5''$ pixels, in
the remaining reduction steps.)
The main effect of this multiplication is to reduce the size of the
image in celestial coordinates, and with it the distance
between each source and the center of the image. 
We bin the data in bins of $5''\times5''$.
We do a first source
detection by comparing the counts in a box with those in the surrounding
ring. The detection box is moved across the image.
The sources thus detected are excised from the image, and a spline
fit is made to the spatial distribution of the remaining photons. 
This fit is used as a background map
for a second source detection pass, in which the excess of photons above
the background is determined at each position.
At the positions of all sources found in the two passes, we subsequently
apply a maximum-likelihood algorithm which compares 
(a Gaussian approximation of) the point spread function of the 
XRT-HRI combination with the distribution of the individual photons,
to determine the probability that a source is detected.
We retain all sources with a likelihood parameter $L>10$ (see Cruddace et
al.\ 1988). \nocite{chs88}

\begin{figure}
   \centerline{\psfig{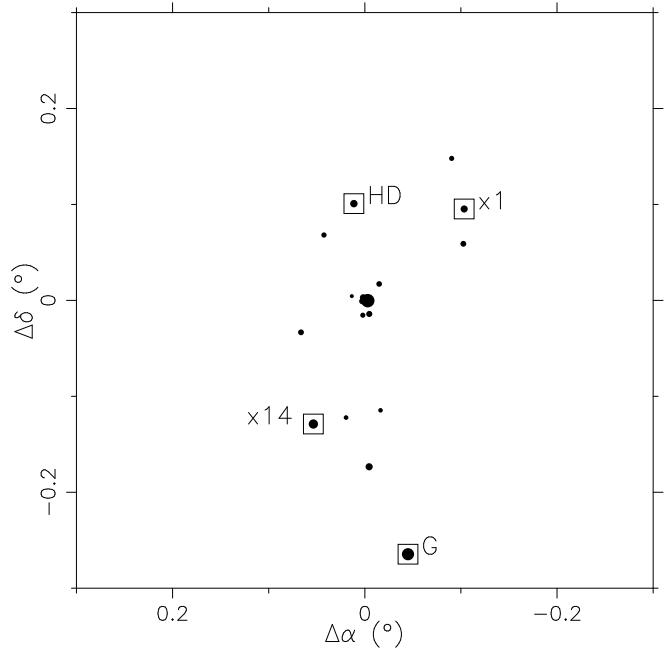}}
   \caption[]{The positions of the sources detected in our HRI observations
of \tuc. Each source is indicated with a $\bullet$, the size of which is
proportional to the logarithm of the observed countrate. 
Positions are with respect to the cluster center as determined by
Guhathakurta et al.\ (1992): 
$\alpha_c(2000) = 00^h24^m5.^s83$, $\delta_c(2000)=-72^{\circ}4'51.''4$.
The sources marked by squares are those used in co-aligning the separate 
observations; HD ($=$\x12) and G ($=$\x3) label the sources identified 
with HD~2072 and a galaxy, respectively,
which have been used to determine the ('bore sight') correction applied
to the X-ray positions to bring them to an optical coordinate frame.
\label{fwhole}}
\end{figure}
\nocite{gysb92}

By comparison of the source lists of the separate observations, we find that
four sources, listed as \x1, \x3, \x12\ and \x14\ in Paper~1 and in
Table~\ref{tb}, are detected in each of the observations b,c,d,e;
\x3, \x12\ and \x14\ are detected in observation f; and \x3\ and \x14\
in observation~a.
By weighted averaging of the position shifts of these sources,
we determine the offsets between the celestial coordinate frame 
of observation~c and that of each of the other observations.
The offsets between each observation and observation c are
listed in Table~\ref{ta}, rounded off to the nearest 
$0.5''$, and are well within the claimed positional
accuracy of the HRI ($5''$).
The offset between observations a and b is large, which is why we
separate the 1992 April and May data.

The offsets listed in Table~\ref{ta}
are applied to bring all HRI observations to the
X-ray coordinate frame of observation~c. After this, all images are added.
The exposure of the added image is 58,820$\,$s.
We repeat the source detection procedure explained above for the
total image, to find the sources listed in Table~\ref{tb} and
shown in Fig.~\ref{fwhole}.

\subsection{Conversion to the optical coordinate frame}

In Paper~1 we suggest HD$\,$2072 as identification for \x12;
Geffert et al.\ (1997) suggest a galaxy at 
$\alpha(2000) = 00^h23^m$ $30.^s8$, $\delta(2000)=-72^{\circ}20'44.''0$
as identification for \x3, \nocite{gak97}
and give the position for HD$\,$2072 as
$\alpha(2000) = 00^h24^m$ $14.^s5$, $\delta(2000)=-71^{\circ}58'49.''1$.
With a scale of $0.5''$ per pixel either of the two suggested identifications
could be accepted, but not both.
With the reduced scale of the X-ray image (see Sect.\ 2.1) we are now
able to accept both identifications simultaneously, \x12\ with  HD$\,$2072 
and \x3\ with the galaxy.

The nominal X-ray positions of \x12\ and \x3\ were at $+0.11^s,+0.6''$ and
$+0.14^s,0.0''$, respectively, from the optical positions, well
within the expected absolute accuracy of the X-ray positions.
Taking into account the more accurate position of source \x12\ we
determine the overall shift between the X-ray frame of observation c
and the J2000 coordinate frame as $-0.12^s(=-0.55''),-0.45''$.
This correction has been applied to obtain
the J2000 positions of the X-ray sources listed in Table~\ref{tb}

Combining the errors of about $1''$ of the location of the two X-ray
sources in the X-ray frame with the error of about $1''$ in
determining the shift between nominal X-ray positions and J2000, we
estimate that the overall accuracy in aligning the X-ray coordinate
frame to J2000 is better than $2''$.

\begin{figure}
   \centerline{\psfig{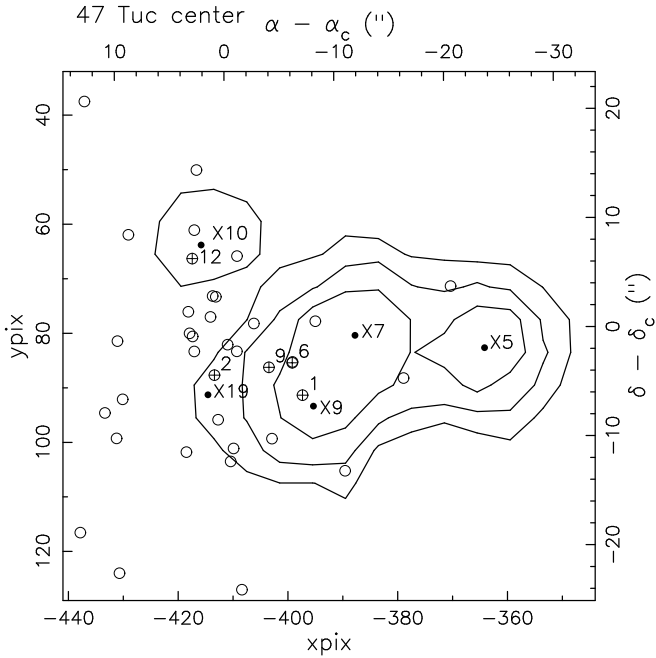}}
   \caption[]{X-ray image of the core of \tuc. The solid lines show
contour levels, determined on the basis of $3''\times3''$ bins,
at 0.4, 0.16 and 0.064 times the maximum observed. $\bullet$-s indicate
the locations of the sources as determined with a multi-source algorithm.
The lower and left axes refer to the detector coordinates; the upper and
right axes give the distance with respect to the optical center of 
\tuc\ as given by Guhathakurta et al. (1992). The uncertainty between the
two coordinate systems is $\ltap 2''$.
Circles indicate locations of the optical variable and blue stars in the
core of \tuc, listed in Table~3 of Geffert et al.\ (1997).
Some stars discussed in the text are inscribed with $+$, and labelled 1,2 for
the blue objects \vv1 and \vv2, respectively; 6,9 for the variables
\ako6 and \ako9, respectively; and 12 for variable 12 of Edmonds et 
al.\ (1996). The Einstein X-ray source (Grindlay et al. 1984)
is not indicated, but coincides closely with \vv1.
\label{fcore}}
\end{figure}
\nocite{egg+96}\nocite{ghs+84}

\subsection{Multiple source detection and variability}

The standard analysis cannot disentangle sources that overlap.
Figure~\ref{fcore} shows that we have such a situation in the
central region of 47 Tucanae.
We therefore apply a multi-object algorithm to determine the
number and positions of the sources in this region.
This algorithm is described in Paper~1: briefly, it 
compares the predicted photon numbers with the observed ones
for a binned data set, for an assumed point spread function.
We use a bin size of $3''\times3''$.
We vary the number of sources, and find that five sources,
listed in Table~\ref{tb}, are required to explain the observation.
Initial positions for the five sources were set by hand, on the
basis of the contour plot shown in Fig.~\ref{fcore}; the best
positions of these sources, and their countrates, were determined
with the multi-source algorithm. The fit for five sources is significantly
better than a four-source fit in which source \x19 is omitted.

\begin{figure*}
   \centerline{\psfig{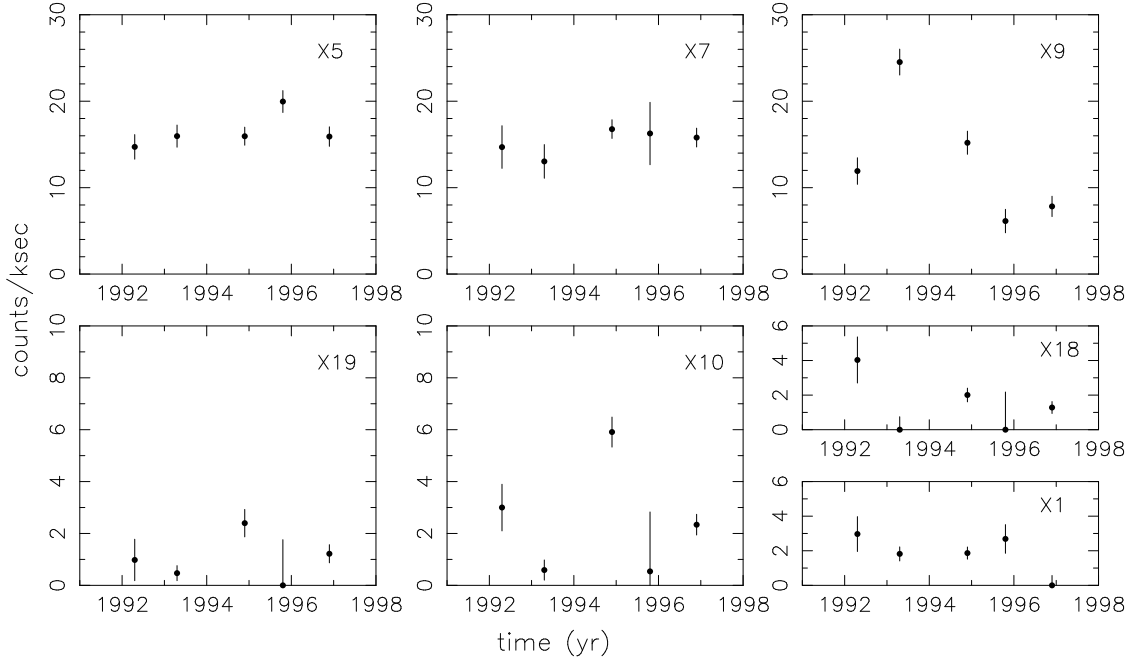}}
   \caption[]{Lightcurves of the five X-ray sources in the core
of \tuc, and of two variable sources probably not related to the
cluster, \x1 and \x18. For non-detections of \x1 and \x18, 
an upper limit of 10 counts has been assumed; the 1992 points for these
sources refer to the May observation only. Amongst the sources 
in the core, \x9 and \x10 are significantly variable.
\label{fvar}}
\end{figure*}

\begin{table}
\caption[o]{Sources detected in the total image of 47~Tuc. 
The exposure time is 58820$\,$s. 
The sources are ordered with respect to their right ascension. 
Sources \x1 to \x15 correspond to the sources in Hasinger et al.\ 1994;
sources \x16 to \x21 are new.
For each source we give the position, the statistical uncertainty
in the position, the detected total number of counts, and for 
sources within 2$'$ from the cluster center and therefore
probably associated with \tuc\ the X-ray luminosity in the 
$0.5-2.5\,$keV band. Errors for source positions found with the multi-source
fit program, marked with $*$, are mutually dependent, and hard to
quantify. We estimate that they are $\sim 3''$ for \x19 and $2''$
for the others.
In addition to the statistical uncertainty in the individual
positions, the source positions are subject to a possible systematic
uncertainty, which we estimate to be less than $2''$.
\label{tb}}
\begin{tabular}{rr@{\hspace{0.1cm}}r@{\hspace{0.1cm}}rr@{\hspace{0.1cm}}r@{\hspace{0.1cm}}rrr@{\hspace{0.0cm}}rr}
 X &
 \multicolumn{3}{c}{$\alpha$(2000)} &
 \multicolumn{3}{c}{$\delta$(2000)} & $\Delta$ & cts & & $L_{\rm x}$(erg/s) \\
 & & & & & & & ($''$) & & & ($\times 10^{32}$) \\
1 &   0 &  22 &  45.38 & -71 &  59 &   8.8 & 0.7 &   89$\pm$ &  11 \\
2 &   0 &  22 &  46.07 & -72 &   1 &  19.8 & 1.0 &   52$\pm$ &   9 \\
16 &  0 &  22 &  55.48 & -71 &  55 &  58.9 & 2.1 &   39$\pm$ &   9 \\
3 &   0 &  23 &  30.82 & -72 &  20 &  44.5 & 1.0 &  720$\pm$ &  32 \\
17 & 0 &  23 & 52.98 & -72 &  11 &  44.3 & 1.4 &   31$\pm$ &   7 \\
4 &   0 &  23 &  54.06 & -72 &   3 &  50.3 & 1.4 &   49$\pm$ &   9 & 0.6 \\
5 &   0 &  24 &   0.93 & -72 &  4 & 53.5 & $*$ & 815.3$\pm$ & 43 & 9.3 \\ 
6 &   0 &  24 &   2.13 & -72 &   5 &  42.7 & 0.8 &   59$\pm$ &   9 & 0.7 \\
18 & 0 &  24 &  2.16 & -72 &  15 &  16.1 & 1.5 &   91$\pm$ &  12 \\
7 &   0 &  24 &   3.49 & -72 &  4 & 52.4 & $*$ & 1098.2$\pm$ & 61 & 12.5 \\
9 & 0 & 24 &  4.31 & -72 &  4 & 58.9 & $*$ & 696.4$\pm$ & 52 & 8.0 \\
19 & 0 &  24 &  6.39 & -72 &  4 & 57.8 & $*$ & 115.9$\pm$ & 15 & 1.3 \\
10 & 0 & 24 & 6.52 & -72 &  4 & 44.1 & $*$ & 178.8$\pm$ & 19 & 2.0 \\
11 &   0 &  24 &   7.34 & -72 &   5 &  47.5 & 1.5 &   39$\pm$ &   8 & 0.4 \\
12 &   0 &  24 &  14.49 & -71 &  58 &  49.0 & 0.6 &  102$\pm$ &  11 \\
13 &   0 &  24 &  16.17 & -72 &   4 &  36.0 & 1.0 &   24$\pm$ &   6 & 0.3 \\
20 & 0 &  24 & 20.80 & -72 &  12 &  11.8 & 1.5 &   32$\pm$ &   7 \\
21 &  0 &  24 & 38.76 & -72 &   0 &  46.5 & 0.9 &   41$\pm$ &   8 \\
14 &   0 &  24 &  47.23 & -72 &  12 &  35.8 & 0.6 &  221$\pm$ &  16 \\
15 &   0 &  24 &  57.20 & -72 &   6 &  52.0 & 1.2 &   52$\pm$ &   9 \\
\end{tabular}
\end{table}

Finally, we investigate the source variability. 
For the sources outside of the core, we compare the countrates determined
with the standard analysis of the individual observations b-f, as described
above. If we allow that a source
with an expected number of counts of $\ltap 10$ can escape detection,
we find that only two significant non-detections remain (see Fig.~\ref{fvar}).
Source \x1 is not detected in the November 1996 observation, whereas
$27$ counts are expected.
Source \x18 is not detected in the April 1993 observation, whereas
$20$ counts are expected.
We find no evidence for $>2\sigma$ excess brightness in individual 
observations.

For the sources in the core, we fix the
source positions as listed in Table~\ref{tb}, and on these positions
determine the countrates for observations a and b combined, and for
each of the observations c to f, with the multi-source algorithm.
(Observations a and b are too short for separate countrate determinations.) 
The resulting individual countrates are shown in Fig.~\ref{fvar}.
Sources \x9 and \x10 are highly variable, the other sources
are not significantly variable.

For conversion of the countrates to luminosities for the sources associated
with \tuc, we assume a 1~keV bremsstrahlung spectrum, absorbed by
$\nh=2.4\times10^{20}$ $\cmsq$. 
At a distance of 4.6~kpc, 1 count per kilosecond
in the HRI then corresponds to a luminosity at the source of 
$6.7\times 10^{31}$erg/s.
For the sources that are probably related to \tuc, we use this conversion
to compute the luminosities listed in Table~\ref{tb}.

\section{Results}

Our analysis leads to the detection of 20 sources in the direction of \tuc.
With the exception of \x8, all sources \x1-15 discussed in Paper~1 are detected
again; six new sources \x16-21 have been added, of which \x19 is in the core.
Source \x8 was an artefact of the large relative shift between the
April and May 1992 X-ray coordinate systems; now that we apply bore sight
corrections to these observations separately, source \x8 is no longer found.
Five sources are detected in the core (Fig.\ 2).
Four sources are detected within 2$'$ from the cluster center, rather more
than expected randomly from the number of sources in the whole image; 
we conclude that these four sources are probably also related to 
\tuc\ (see Fig.\ 1).
Thus nine sources have been detected in the cluster.

\subsection{Inside the core}

In Fig.~\ref{fcore} we show the 5 X-ray sources in the core, together with the
variable and blue stars listed by Geffert et al.\ (1997; their Table 3).
The X-ray sources are plotted in the detector frame, the optical sources
in the J2000 frame. The relative positions of these frames is accurate
to better than $2''$.
As a result, we can be more confident about possible optical counterparts
to the X-ray sources than has been hitherto possible.
\ako6 and \ako9 do not correspond to any of the detected X-ray sources.
As these sources lie in the wings of the point spread functions of the
detected X-ray sources, we cannot exclude that they emit X-ray flux at
a lower level than the faintest detected X-ray source in the core, \x19.
Positional coincidences within the error are found for \vv1 and \x9,
\vv2 and \x19, and entries 29 and 31 from Table~3 in Geffert et al. (1997)
with \x10.

\vv1 was discovered by Paresce et al.\ (1992); according to Shara et 
al.\ (1996) \vv1 is not significantly variable in the Hubble Space
Telescope images. \nocite{sbg+96}
Thus the nature of the source is not clear. It may be a cataclysmic
variable; its magnitudes are also compatible with those of a low-mass
X-ray binary in a quiescent state.
The ultraviolet flux reported by Paresce et al.\ (1992) corresponds
to an $AB_{\nu}$ magnitude, corrected for the reddening towards \tuc,
of $AB_{\nu}\simeq 20.5$.
The position of \vv1 coincides within the accuracy with \x9, and also with the 
position of the X-ray source detected in \tuc\ with Einstein 
(Grindlay et al.\ 1984).
\nocite{pmf92}\nocite{sbg+96}\nocite{ghs+84}
Both the Einstein observations (Auri\`ere et al.\ 1989) and our ROSAT HRI
observations (Fig.~\ref{fvar}) show that the X-ray flux is variable by a 
factor $\sim 3$.

\begin{figure}
   \centerline{\psfig{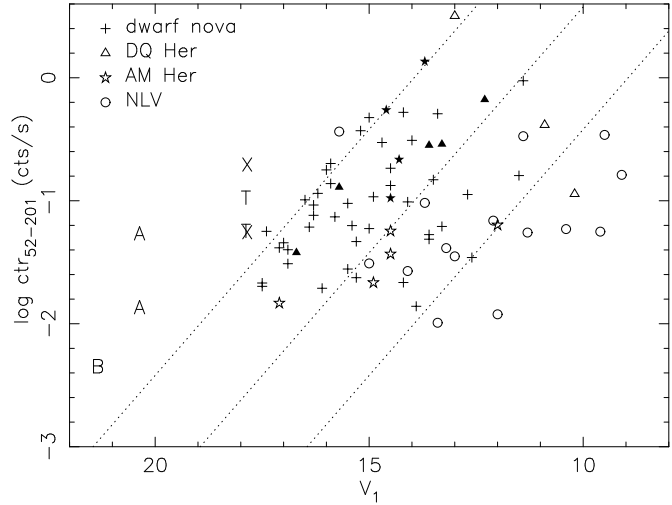}}
   \caption[]{ROSAT PSPC countrates of cataclysmic variables detected in
the ROSAT All Sky Survey, as a function of visual magnitude. The data
are from Verbunt et al.\ (1997), except that we take $V_1=15.9$ for BZ UMa.
(Index 1 indicates the magnitude at which the system is most frequently
found, MAG1 in Ritter 1990.) \nocite{rit90}
The dashed lines correspond to a constant ratio of X-ray to 
ultraviolet/optical flux. Different symbols indicate dwarf novae,
AM~Her and DQ~Her type systems, and other cataclysmic variables
('nova like variables'); solid symbols indicate systems detected
first in X-rays, and subsequently identified with cataclysmic variables
('X-ray selected systems').
We also show with A and B the estimated positions in this diagram for 
\x9 (at low and high observed countrate) and \x19,
if we assume that these may be identified with \vv1 and \vv2, respectively.
T and X indicate the locations of Aql X-1 and Cen X-4, respectively,
in two observations at quiescence, corrected for interstellar absorption
(optical data from Van Paradijs 1995; X-ray data from Verbunt et al.\ 1994,
Campana et al.\ 1997).
\label{fcvs}}
\end{figure}

To investigate the nature of \x9 and \vv1 we compare its visual magnitude
and X-ray flux with those of cataclysmic variables from the Rosat All Sky
Survey in Fig.~\ref{fcvs}.
We estimate the countrate in channels 50--201 of the PSPC by multiplying 
the HRI countrate listed in Table~\ref{tb} with a factor 2; and assume
that the visual magnitude equals the $AB_{\nu}$ ultraviolet magnitude
(as is often, but not always, the case for cataclysmic variables; see
e.g.\ Verbunt 1987). \nocite{ver87b}
We see in Fig.~\ref{fcvs} that the ratio of the X-ray flux to the optical
flux is rather high for a cataclysmic variable; even at the lowest measured
X-ray flux of \x9, the X-ray to optical flux ratio is higher than that
of any cataclysmic variable detected in the Rosat All Sky Survey.
If \vv1 is identical to \x9, we suggest therefore
that it is a low-mass X-ray binary in a low state, i.e.\ the accreting
object is a neutron star. Two soft X-ray transients in their
quiescent state, Cen X-4 and Aql X-1, are also shown in Fig.~\ref{fcvs};
their X-ray to optical flux ratio is similar to those of \vv1/\x9.

\vv2 is a blue variable detected twice at a high level about 4 magnitudes above
its quiescent level (Paresce \& De Marchi 1994, Shara et al.\ 1996).
In the ultraviolet, its quiescent $AB_{\nu}$ magnitude is 
$AB_{\nu}\simeq 21.5$. \nocite{pm94}
\nocite{pm94} Its position coincides within the accuracy with \x19. 
Assuming again that the PSPC countrate is about twice the HRI countrate,
and that the $V\simeq AB_{\nu}$, we show the location of 
\vv2 and \x19 in Fig.~\ref{fcvs}.
If \vv2 and \x19 are identical, than the ratio of X-ray to optical flux
is higher also for this system than for any cataclysmic variable detected
in the ROSAT All Sky Survey, and similar to those of Cen X-4 and Aql X-1 in
quiescence.

\nocite{vbrp97}\nocite{cmsc97}\nocite{vbj+94}\nocite{vpa95}

Entries 29 and 31 of Table~3 in Geffert et al.\ (1997) are nos.\ 1030 and 1286
of De Marchi et al.\ (1993), and both are blue stragglers. \nocite{mpf93}
Entry 31 of Geffert et al.\ (1997) corresponds to binary no.\ 12 in
Edmonds et al.\ (1996). The binary period is 0.69 days, or possibly 
1.38 days; the nature of the binary is not clear.
Further study of these systems is required before one of them 
can be identified with \x10.

A third blue variable has been found in the core of \tuc\ by Shara et 
al.\ (1996). The nature of this variable, \vv3, is not known.
From Fig.~3c in Shara et al.\ (1996) we estimate that \vv3 is about
9.4$''$ North of \vv2; and about 0.8$''$ East. This is close to \x10,
but too far for \vv3 to be a candidate for identification with \x10.

The two brightest constant sources \x5 and \x7 cannot be identified
with any of the blue and variable stars known in the core of \tuc.
It is remarkable that the Einstein satellite did not detect these
sources, but only the variable source \x9.
This suggests that \x9 was significantly brighter during the
Einstein observations than \x5 or \x7; inspection of Fig.~\ref{fvar}
suggests that \x9 occasionally may indeed get bright enough
to outshine \x5 and \x7.

\subsection{Near the core; extended emission}

\begin{figure}
   \centerline{\psfig{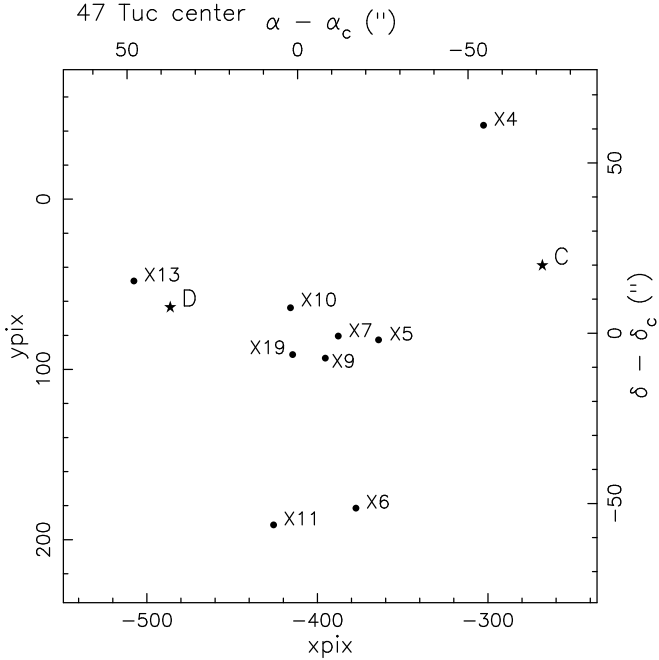}}
   \caption[]{Location of the X-ray sources in and near the core
of \tuc, compared with the two known positions of radio pulsars
\tuc~C and \tuc~D.
\label{flcore}}
\end{figure}

Figure~\ref{flcore} shows that there are four sources outside but near the
core of \tuc.
The source density of the whole image is such, that these four sources
all probably are members of \tuc.
In Fig.~\ref{flcore} we show the positions of the central sources of
our HRI observations together with the two radio pulsars whose positions
are known (to about 0.01$''$): \tuc~C and \tuc~D (Robinson et al.\ 1995). 
\nocite{rlm+95}
Neither pulsar is detected in X-rays.
No less than 11 radio pulsars have been detected in \tuc; determination
of the positions of pulsars \tuc~E to N is awaited to see whether any
of them coincides with an X-ray source.

We investigate the existence of possible extended emission in two
ways.
First, we determine a total number of 3780 counts 
in a $100''\times100''$ region centered on the central sources,
and then subtract the 3200 counts assigned by our multi-source fit
to individually detected sources in this region.
Of the excess of about 580 counts, some 230 are
expected in the wings of the point spread functions of the
individual sources, leaving about 350 counts.
A similar excess is found by comparing the radial distribution
of the detected counts with that of the point spread function
between 20$''$ and 60$''$ from the X-ray center of the cluster.
We conclude that there is extended emission in and/or near the
core of \tuc, with a countrate of $\sim 6\,$cts/ksec, which corresponds
to an X-ray luminosity at the distance of \tuc\ of roughly
$L_{\rm 0.5-2.5 keV}\sim 4\times10^{32}$erg/s.
We cannot decide on the basis of our data whether this excess is
due to one or two individual faint sources, or to a larger number
of even fainter ones.

\subsection{Sources not related to \tuc}

In the wider field of view shown in Fig.~\ref{fwhole}, we have
optically identified just two sources.  The identification of \x12 with
HD$\,$2072 in our view is fairly secure, for the reasons given in
Paper~1: HD$\,$2072 has visual magnitude $m_{\rm V}\simeq9.1$, and the
number of objects this bright in the optical is so small as to make
chance coincidence unlikely. 
With $M_{\rm V}=5.2$ for a G5 main-sequence star, we derive a distance
of about 55~pc, and an X-ray luminosity for
\x12 of $L_{\rm 0.5-2.5 keV}\sim 1.7\times10^{28}$erg/s, quite
reasonable for a G$\,$5 star.
(The smaller distance and luminosity correct the values given in Paper~1.)

The identification of \x3 with a galaxy is somewhat less secure,
in the sense that faint galaxies are sufficiently common to make
a chance coincidence possible.
Because the X-ray position of \x12 is more accurate than that of
\x3, the shift between the optical and X-ray frame is determined mainly
by the identification of \x12 with HD$\,$2072, and thus is not
affected much by the correctness of the identification of
\x3 with the galaxy.

\section{Discussion}

We have detected five X-ray sources in the core of \tuc, and noted
possible optical identifications for three of them.

Before we discuss the possible nature of these sources, we adress the
question how confident we can be about the identifications.
To do this, consider an area of $20''\times20''$, centered on the
cluster center according to Guhathakurta et al.\ (1992).
From Fig.~\ref{fcore} we learn that this area contains three X-ray
sources and 22 blue or variable stars. (Note that entry 8 of Table~3
in Geffert et al.\ (1997) almost coincides with \ako6.)
If we suggest identification for each blue object lying in a $4''\times4''$
box centered on an X-ray source, then the X-ray sources cover 12\%\ of the 
search area, and we have 22 trials for probability 0.12.
The probabilities of finding 0, 1, 2 or 3 identifications are
6, 18, 26 and 23 \%, respectively.
We conclude that the probability that all suggested identifications
are accidential is quite high.

It may be argued that the suggested identifications are special also
optically. 
If we consider the three objects \vv1, \vv2 and \vv3 only, we have
3 trials for probability 0.12, and the probability of finding
0, 1 or 2 identifications are 68, 28 and 4 \%.
Even for this limited set, the probability that both identifications
of \vv1 with \x9 and \vv2 with \x19 are due to chance is non-negligible.
For the moment we conclude that our suggested identifications are
possible, but not secure.

If we assume that \vv1 and \vv2 may be identified with \x9 and \x19,
respectively, we learn from Fig.~\ref{fcvs} that their ratio of
X-ray to optical flux is rather high if they are cataclysmic
variables, but as expected for soft X-ray transients in quiescence.
The X-ray countrates of the cataclysmic variables in Fig.~\ref{fcvs}
have not been corrected for interstellar absorption; the correction is
expected to be small for most systems, but not necessarily for all.
For typical X-ray spectra of cataclysmic variables, the visual flux
is affected more strongly by interstellar absorption than the X-ray
countrate, and thus it is not expected that correction for absorption
will increase the ratio of X-ray to optical flux for cataclysmic
variables.
We conclude that Fig.~\ref{fcvs} provides another illustration of the 
argument originally made by Verbunt et al. (1984) \nocite{vpe84} that some
of the dim X-ray sources in the cores of globular clusters are
too bright to be cataclysmic variables.

The X-ray flux of \x9 is variable; that of \x19 may or may not be variable.
The range of variability in \x9 is not unprecedented in soft X-ray transients
in quiescence: the variations in the flux of Cen X-4 in quiescence, 
reported by Campana et al.\ (1997) and shown in Fig.~\ref{fcvs}, is of
a similar magnitude as that observed in \x9.
Such variations in a quiescent soft X-ray transient are not expected to
be accompanied by detectable optical variations, and thus the absence
of optical variation in \vv1\ need not be in conflict with the suggested 
identification.
It may be noted that similar variations in the X-ray flux without accompanying
variations in the optical are probably also possible in cataclysmic variables.
For example, the dwarf nova VW~Hyi was brighter in quiescence when
observed with ROSAT in Nov 1990 than when observed with EXOSAT several
years earlier (Wheatley et al.\ 1996, their Fig.~7).\nocite{wvb+96}

\vv2 has been detected at a level about 4 magnitudes above its quiescent
level twice; this magnitude difference is more indicative of a dwarf
nova than of a soft X-ray transient, as noted by Paresce \&\ De Marchi
(1994) and by Shara et al.\ (1996).

The two constant X-ray sources \x5 and \x7 in the core of \tuc\ have no 
suggested optical counterparts.
The level and the constancy of their X-ray fluxes are compatible with
them being radio pulsars.
For example, PSR$\,$B$\,1821-24$ in globular cluster M$\,$28 and
PSR$\,$J$\,0218+4232$, at comparable distances as \tuc\ (5.5 and $>$5.7 kpc,
respectively compared to 4.6 kpc for \tuc), have ROSAT PSPC countrates
of the same order of magnitude as \x5 and \x7.
Whether \x5 or \x7, or any of the four X-ray sources just outside the
core, can be identified with any of the 11 radio pulsars in \tuc\ awaits
further study of the radio pulsars, in particular determination of their
positions, and of their period derivatives (so that the X-ray data
can be folded on a known period).
More accurate pinpointing of the X-ray positions will be possible with
AXAF.
Considering the large numbers of potential optical counterparts, 
optical or ultraviolet monitoring of the inner region of \tuc\
simultaneous with the X-ray observations would be very useful, as
detection of simultaneous X-ray and optical variability would
strenghten any identification based on positional coincidence only.

To summarize, we find possible optical counterparts for three of the
five X-ray sources in the core of \tuc, but note that all could be
chance positional coincidences. The X-ray luminosities of \x5, \x7
and \x9 are rather high for these to be cataclysmic variables, but
compatible with soft X-ray transients in quiescence. \x9 is a variable
X-ray source, and its X-ray to optical flux ratio suggests that it is
a soft X-ray transient, hitherto always observed in quiescence.  The
steadier sources \x5 and \x7 may be either soft X-ray transients or
recycled radio pulsars.  The sources \x19 in the core,
and \x4, \x6, \x11 and \x13, outside
but near the core, have X-ray luminosities $L_{\rm X}<10^{32}$erg/s,
compatible with them being soft X-ray transients, cataclysmic
variables, or recycled radio pulsars. 
If \vv2 is indeed a cataclysmic variable, it is probably the best
candidate counterpart hitherto suggested for an X-ray source in \tuc.

\acknowledgements{We thank M.\ Geffert for information before
publication on astrometry of objects near \tuc.
We thank M.\ Auri\`ere, A.\ Cool, G.\ De Marchi,
G.\ Djorgovski, J.\ Grindlay, P.\ Guhathakurta, H.\ Johnston, L.\ Koch,
M.\ Livio, B.\ Margon, G.\ Meylan and F.\ Paresce for supporting our
target of opportunity request for a ROSAT observation quasi simultaneous
with a Hubble Space Telescope observation (reported in Minniti et 
al.\ 1997) in October 1995; and especially J.\ Schmitt for scheduling 
this observation in spite of a solar eclipse on the day in question.}


\end{document}